%% file: main.tex
\documentclass[12pt,preprint]{aastex}
\usepackage{natbib}
\def\be{\begin{equation}}
\def\ee{\end{equation}}

\def\m{~$\mu$m}
\def\R25{$R_{25}$}

\def\i    {{\it i}}

\def\2MASS{{\it 2MASS}}

\def\SDSS{{\it SDSS}}

\begin {document}

\title{RADIAL STAR FORMATION HISTORIES IN 32 NEARBY GALAXIES}
\shorttitle{Age Gradients}

\input authors.tex
\input abstract.tex
\input paper.tex
\input acknowledgements.tex

\input natbib.bbl
\input tables.tex
\input figs.tex
\end{document}

%% file: authors.tex
\author {
Daniel~A. Dale\altaffilmark{1},
Kristin R. Anderson\altaffilmark{2}, 
Louis~M. Bran\altaffilmark{3}, 
Isaiah~S. Cox\altaffilmark{4}, 
Carolyn~L. Drake\altaffilmark{5}, 
Nathan~J. Lee\altaffilmark{1}, 
Jacob~D. Pilawa\altaffilmark{6}, 
F.~Alexander Slane\altaffilmark{1}, 
Susana Soto\altaffilmark{7}, 
Emily~I. Jensen\altaffilmark{1},
Jessica~S. Sutter\altaffilmark{1},
Jordan~A. Turner\altaffilmark{1}, and
Henry~A. Kobulnicky\altaffilmark{1}
}
\altaffiltext{1}{Department of Physics \& Astronomy, University of Wyoming, Laramie WY; ddale@uwyo.edu}
\altaffiltext{2}{Department of Physics \& Astronomy, California State University, Long Beach, CA}
\altaffiltext{3}{Department of Mathematics \& Applied Physics, California State University Channel Islands, Camarillo, CA}
\altaffiltext{4}{Department of Physics \& Astronomy, East Tennessee State University, Johnson City, TN}
\altaffiltext{5}{Department of Astronomy, Whitman College, Walla Walla, WA}
\altaffiltext{6}{Department of Physics \& Astronomy, Colgate University, Hamilton, NY}
\altaffiltext{7}{Department of Physics \& Astronomy, University of California, Irvine, CA}

%% file: abstract.tex
\begin {abstract}
The spatially resolved star formation histories are studied for 32 normal star-forming galaxies drawn from the the {\it Spitzer} Extended Disk Galaxy Exploration Science survey.  At surface brightness sensitivities fainter than 28~mag~arcsec$^{-2}$, the new optical photometry is deep enough to complement archival ultraviolet and infrared imaging and to explore the properties of the emission well beyond the traditional optical extents of these nearby galaxies.  Fits to the spectral energy distributions using a delayed star formation history model indicate a subtle but interesting average radial trend for the spiral galaxies: the inner stellar systems decrease in age with increasing radius, consistent with inside-out disk formation, but the trend reverses in the outermost regions with the stellar age nearly as old as the innermost stars.  These results suggest an old stellar outer disk population formed through radial migration and/or the cumulative history of minor mergers and accretions of satellite dwarf galaxies.  The subset of S0 galaxies studied here show the opposite trend compared to what is inferred for spirals: characteristic stellar ages that are increasingly older with radius for the inner portions of the galaxies, and increasingly younger stellar ages for the outer portions. This result suggests that either S0 galaxies are not well modeled by a delayed-$\tau$ model, and/or that S0 galaxies have a more complicated formation history than spiral galaxies.
\end {abstract}


%% file: paper.tex
\section{Introduction}
\label{sec:intro}

Integral field spectroscopy surveys have pushed forward our understanding of the spatially resolved star formation histories of nearby galaxies ($z \lesssim 0.1$).  The Calar Alto Legacy Integral Field Area \citep[][CALIFA;]{sanchez12}, the Herschel ATLAS3D survey \citep{cappellari11}, the Sydney-AAO Multi-object Integral field spectrograph survey \citep[][SAMI;]{bryant15}, and the Mapping Nearby Galaxies at APO survey \citep[MANGA;][]{bundy15} are carrying out integral field surveys of hundreds to thousands of galaxies in the local universe.  Other approaches besides integral field surveys can lead to constraints on spatially resolved galaxy star formation histories, including high-resolution imaging combined with color-magnitude diagram analysis for resolved stellar populations in the closest galaxies \citep{weisz13,meschin14} and comparatively coarse-resolution panchromatic broadband imaging surveys coupled with SED fitting \citep{mejianarvaez17,williams18,smith18}.  Many of these extragalactic studies point to the so-called ``inside-out'' growth of galaxy disks.  In such a scenario disks may grow through star formation triggered by the accretion of pristine intergalactic gas funneled from the cosmic web, or through mergers and the accretion of small satellites \citep[e.g.,][]{abadi03,governato04,robertson06,governato07}.  Observational markers of galaxies with a history of inside-out disk formation include negative metal abundance gradients and ultraviolet/optical colors that are increasingly blue as a function of galactocentric radius \citep{larson76,ryder94,dejong96,avilareese00,macarthur04,munozmateos07,wang11,pilkington12,barnes14,dsouza14,rodriguezbaras18,mondal19}.  Star formation history models fitted to spatially resolved multi-wavelength datasets also suggest inside-out growth for spiral galaxies, with younger stellar populations inferred further out in the disks \citep[e.g.,][]{brown08,williams09,gogarten10,barker11,monachesi13,garciabenito17,lopezfernandez18}.  A handful of studies suggest evidence for an outside-in formation process, but only for low-mass dwarf galaxies \citep[e.g.,][]{gallart08,zhang12,meschin14,pan15,sacchi18}.  

The structure and physical characteristics of the outermost portions of galaxies, beyond the ``disk proper'', also provide key clues to their formation histories \citep[e.g.,][]{ferguson98,thilker07}.  For example, a significant fraction of spiral galaxies exhibit ``downbending'', where the optical surface brightness profile more steeply drops in the galaxy outskirts compared to that measured for the main disk \citep{gutierrez11}.  Furthermore, this downbending phenomenon (or its upbending converse), appears to depend on galaxy morphology: multiple studies, each spanning dozens of galaxies, show that downbending (upbending) occurs more frequently for later-type (earlier-type) spirals \citep{pohlen06,martinnavarro12,staudaher19}.  Upbending in the furthest reaches of a surface brightness profile is particularly intriguing since it may reflect the build-up of stellar haloes through multiple prior mergers and accretion events \citep{abadi06,read06,purcell07,cooper13}.  An alternative hypothesis for upbending in a galaxy's periphery relies on simulations exhibiting ``radial migration'', whereby internal dynamical processes redistribute stars originally formed near the center \citep{sellwood02,roskar08,radburnsmith12}. 

Probing the outermost regions of nearby galaxies requires both wide field imaging and deep integrations.  The Extended Disk Galaxy Exploration Science (EDGES) survey was a 1000$+$~hour Warm Spitzer program to probe the outermost extent of the old stellar populations in 92 nearby galaxies ($z < 0.004$).  The program was designed to carry out wide-field imaging out to five times the traditional optical size and down to the sensitivity limits of Spitzer at 3.6 and 4.5\m\ ($\sim 0.4$~kJy or 29~AB~mag~arcsec$^{-2}$).  Details on the full survey including its observational strategies, the data processing, and the analysis of the surface brightness profiles, can be found in \cite{staudaher19}.  In \cite{dale16} we presented the results from an analysis of the radial star formation history profiles for a subset of 15 EDGES galaxies.  The analysis was buttressed by deep multi-wavelength photometry spanning ultraviolet, optical, and infrared wavelengths.  Here we extend the work presented in \cite{dale16} by analyzing similar multi-wavelength photometry for an additional 17 EDGES galaxies; the total sample studied here comprises 32 EDGES galaxies.  Energy-balanced spectral energy distribution (SED) fits are carried out with the CIGALE software \citep{boquien19} with the primary aim of extracting stellar mass-weighted ages as a function of galactocentric radius.  Section~\ref{sec:sample} presents the galaxy sample, Section~\ref{sec:data} reviews the new and archival data compiled for this analysis along with a brief overview of the data processing, Section~\ref{sec:analysis} explains the analysis including the SED fitting, Section~\ref{sec:results} presents the main results, and Section~\ref{sec:summary} provides a summary and brief discussion.

\section{Galaxy Sample}
\label{sec:sample}

Table~\ref{tab:sample} provides the list of 32 EDGES galaxies studied here.  The galaxies represent a subset of EDGES galaxies observable during the summer from the Wyoming Infrared Observatory (WIRO) and served as the foci of the work carried out by Wyoming REU (Research Experience for Undergraduates) interns in 2014 and 2018.  Priority was given to targets with deep ancillary ultraviolet and infrared data (see \S~\ref{sec:data}).  The overall EDGES sample contains 92 nearby galaxies spanning a range of morphology, luminosity, and environment, for galaxies at high Galactic latitudes $|b|>60$\degr, with apparent magnitudes $m_B<16$, and optical angular diameters $2 \lesssim D(^\prime) \lesssim 13$.  The sample was restricted to higher Galactic latitudes to minimize contamination from foreground Milky Way stars.  Six of the galaxies in this subset of the EDGES sample have S0 morphology, ten are irregulars, and the remaining 16 targets are spiral galaxies.  The 32 galaxies studied here have distances between $\sim 3$ and 22~Mpc with a median value of $\sim 9$~Mpc.  The only close pairs, defined by \cite{depropis07} as having projected separations $<$~20~kpc and recessional velocity differences $<$~500~km~s$^{-1}$, are NGC~4485/NGC~4490 and NGC~4618/NGC~4625.  Figure~\ref{fig:color_luminosity} demonstrates the subsample's range of optical $g-r$ colors and $r$ luminosities; a few galaxies appear in the red sequence and green valley near the top, but most of the subsample spans the blue cloud.  Figure~\ref{fig:sample} displays the distributions of four different physical properties as a function of optical morphology.  The global $g-r$ colors are taken directly from our observations, whereas the other three properties---age, star formation rate surface density, and stellar mass---are inferred from the SED fitting described in \S~\ref{sec:sed}.

\section{Data}
\label{sec:data}

\subsection{{\it Spitzer Space Telescope} 3.6\m\ Data}
\cite{staudaher19} describe in detail how the near-infrared mosaics were constructed for EDGES galaxies.  The mosaics are quite large, tracing the 3.6\m\ emssion out to at least five times the optical radius $a_{25}$\footnote{$a_{25}$ is defined as the length of the semi-major axis for the $B$-band isophote at 25~mag~arcsec$^{-2}$ \citep{devaucouleurs91}.}.  Additionally, the EDGES near-infrared mosaics probe to fainter surface brightness levels than other {\it Spitzer}/IRAC imaging campaigns of nearby galaxies.  The EDGES program achieved 1800~s integration per position on the sky, 7.5 times longer than the integrations for the SINGS \citep{kennicutt03}, LVL \citep{dale09b}, and S$^4$G \citep{sheth10} surveys, and 12--30 times longer than the IRAC GTO project \citep{pahre04}.  We reach a 1$\sigma$ per pixel sensitivity of 2~kJy~sr$^{-1}$, and averaging over annuli spanning several square arcminutes can reach down to sensitivities below 0.4~kJy~sr$^{-1}$ (fainter than 29~mag~arcsec$^{-2}$~AB), a level necessary for securely detecting the faint outer substructures associated with nearby galaxies \citep{purcell07,krick11,barnes14,staudaher15}.  For comparison, {\it WISE} achieves a 3.4\m\ diffuse sensitivity of at least 1~kJy~sr$^{-1}$ over a $5^\prime$$\times$$5^\prime$ area \citep{wright10,jarrett20}. 

\subsection{Ancillary Ultraviolet and Infrared Data}
Archival ultraviolet and infrared space-based data were gathered from the {\it GALEX} (0.15 and 0.23\m), {\it Spitzer} (8.0 and 24\m), {\it WISE} (12 and 22\m), and {\it Herschel Space Observatory} (70\m) archives.  The ultraviolet observations with GALEX primarily trace continuum emission from massive stars.  The majority of the far- and near-ultraviolet images utilized here arises from integrations longer than 1~ks; we only rely on GALEX All-Sky Imaging Survey data, which were obtained with integrations shorter than 1~ks, for NGC~5523, NGC~5608, and UGC~8303.  The infrared images employed here for the energy-balanced SED fitting (\S~\ref{sec:sed}) trace polycyclic aromatic hydrocarbon emission and the underlying dust grain continuum \citep[e.g.,][]{smith07} in the case of 8.0 and 12\m, or chiefly warm dust in the case of 22, 24, and 70\m\ emission.  \cite{dale16} review the surface brightness sensitivities for these archival data.  These ancillary/archival datasets have native angular resolutions of $\sim 5-6$\arcsec, or they were smoothed to this resolution.

\subsection{New Optical Observations and Data Processing} 
\label{sec:new}
New deep {\it ugr} imaging was obtained for 17 EDGES galaxies on the WIRO 2.3~m telescope with the WIRO DoublePrime camera \citep{findlay16} over the course of the summer of 2018.  The ground-based WIRO data previously obtained for 15 additional EDGES galaxies are described in \cite{dale16}.  DoublePrime is a four-amplifier 4096$\times$4096 camera with $\sim$0\farcs58 pixels, for an overall field of view of 39\arcmin$\times$39\arcmin. For each galaxy and each filter 12 individual 300~s frames were taken.  Individual frames were randomly dithered with small offsets for enhanced pixel sampling.  The typical atmospheric seeing was 1\farcs5--2\farcs0.  Each night a series of zero second bias frames were obtained in addition to a series of twilight sky flats within each filter.

The optical images were processed with standard procedures, including subtraction of a master bias image and removal of pixel-to-pixel sensitivity variations through flatfield corrections.  The 12 dithered 300~s frames for a galaxy taken in one filter were aligned and stacked (summed), resulting in images with integrations equivalent to one hour.  The stacked images are flat to 1\% or better on 20\arcmin\ scales.  The limiting surface brightnesses are $\sim$28~mag~arcsec$^{-2}$~AB.  
The astrometric solutions and flux zeropoints were calibrated using positions and photometry extracted from Sloan Digital Sky Survey \citep[SDSS;][]{york00} imaging on several ($N\gtrsim15$) foreground stars spread across each image stack.  The uncertainties in the zeropoint calibrations were typically 2\%.  Similar to what was done for the imaging at all other wavelengths studied here, foreground stars and background galaxies were removed from each optical image using IRAF/{\tt IMEDIT} and a local sky interpolation.  This editing typically reached down to sources with flux densities of several microJanskys ($\sim21-22$~mag~AB).  

\section{Data Analysis}
\label{sec:analysis}

To facilitate a spatially consistent panchromatic analysis, the higher angular resolution images (WIRO {\it ugr} and {\it Spitzer} 3.6 and 8.0\m) were smoothed to $\sim$6\arcsec\ resolution using a Gaussian profile.  This resolution is comparable to the angular resolutions of the {\it GALEX}, {\it Spitzer} 24\m, {\it WISE} 12 and 22\m, and {\it Herschel} 70\m\ imaging.  A local sky value for each image was estimated and removed via a set of circular apertures located just beyond the outermost reaches of the galaxy emission (see Figure~\ref{fig:aps} and Figure~2 of \citealt{dale16}).

\subsection{Elliptical Photometry}
\label{sec:photometry}

Multi-wavelength photometry was carried out for each galaxy using a series of six concentric elliptical annuli.  The annuli cover semi-major axis $a$ ranges extending to 1.5 times the de Vaucouleurs radius $a_{25}$, namely 0$<$${a} \over {a_{25}}$$<$0.25, 0.25$<$${a} \over {a_{25}}$$<$0.5, 0.5$<$${a} \over {a_{25}}$$<$0.75, 0.75$<$${a} \over {a_{25}}$$<$1, 1$<$${a} \over {a_{25}}$$<$1.25, and 1.25$<$${a} \over {a_{25}}$$<$1.5.  The results presented in \S~\ref{sec:results} are effectively insensitive to the choice of annular widths; similar results are obtained when using annular widths that are narrower or wider by a factor of 1.5.  All annuli for a given galaxy used the same (NED-based) centroids, position angles, and ellipticities.  Photometric uncertainties $\epsilon_{\rm total}$ are computed by summing in quadrature the calibration error and the measurement uncertainty as described in \cite{dale16}.  All fluxes were corrected for Galactic extinction adopting the results of \cite{schlafly11} by assuming $A_V/E(B-V)\approx3.1$ and the reddening curve of \cite{draine03}.

\subsection{SED Fitting}
\label{sec:sed}

Fitting theoretical models to observed spectral energy distributions has become a common technique for estimating the physical properties of galaxies \citep{hunt19}.  We fit the broadband SEDs for our 32 galaxies using the CIGALE software package and its large grid of models \citep{boquien19}.  Though CIGALE allows for the traditional approach of estimating physical parameters via the single best-fit model ($\chi^2$ minimization), we opt for the Bayesian-like approach that weights all the models in the chosen grid (the priors) based on their likelihood values ($\exp^{-\chi^2}$).  The distributions of likelihoods are then used for estimating the physical parameters and their associated uncertainties.  CIGALE also invokes an energy-balanced approach in which the amount of energy that is absorbed by dust in the ultraviolet/optical regime reappears in the infrared as dust emission.  We utilize the stellar and dust emission libraries of \cite{bruzual03} and \cite{dale14}, respectively, the \cite{chabrier03} stellar initial mass function, and a dust attenuation curve ({\tt dustatt\_modified\_starburst}) based on the work of \cite{calzetti00} and \cite{leitherer02}.  The fit parameters, listed in Table~\ref{tab:parameters}, include metallicity $Z$, extinction $E(B-V)$, a power-law exponent $\delta$ that modifies the slope of the attenuation curve, and the $e$-folding decline rate $\tau$ of the galaxy's star formation history.  The main physical outputs are stellar mass $M_*$, stellar mass-weighted age, and star formation rate.  As described in \cite{dale16}, we explored single and double exponential star formation histories, but we have opted to present results using the so-called delayed star formation history (``delayed--$\tau$'') model \citep{lee10,lee11,schaerer13}, i.e.,
\be
SFR(t) \propto \mathcal{A}_0 t e^{-(t-t_0)/\tau}, \;\;\; \mathcal{A}_0(t-t_0<0)=0.
\ee
where the maximum star formation rate occurs at the value of $\tau$ after the onset of star formation: $t-t_0=\tau$.  We prefer the delayed star formation history model since it is a common, simple prescription that relies on a small number of parameters.  Moreover, CIGALE-based simulations show that the delayed star formation model provides superior accuracy in recovering galaxy stellar masses and star formation rates \citep{buat14,ciesla15}.  However, it should be noted that our main conclusions do not change if we opt instead for a single or double exponential star formation history.  We fix $t_0=11$~Gyr since the results of the fits proved to be insensitive to reasonable values of $t_0$.  As a result of fixing the onset of star formation $t_0$, differences in $\tau$ directly correspond to differences in the timing of the peak of the star formation histories.

Representative SED fit results are provided in Figure~\ref{fig:sed} for UGC~7699, the same galaxy portrayed in Figure~\ref{fig:aps} with our approach to the location of photometric apertures.  The fits are systematically low compared to the observations for the GALEX $NUV$ filter and mostly systematically high compared to the $Spitzer$ 8\m\ fluxes, but otherwise the fits for all six annular regions are generally good, with reduced $\chi^2$ values ranging from 0.5 to 1.9.  

Sample-wide, the flux signal-to-noise values naturally are smaller for the outermost annuli.  The median signal-to-noise values, in order of increasing filter wavelength, are [13,11,4,13,13,23,3,3] for the penultimate annulus and [7,3,3,4,5,8,3,3] for the outermost annulus.  There is no minimum recommended signal-to-noise for CIGALE.  In CIGALE, each band is weighted by the inverse variance, and bands with lower signal-to-noise will thus have a lower weight in the computation of the likelihood.  In certain situations this weakens the constraints on some of the physical properties, and this will be reflected in larger uncertainties.  For example, a lower signal-to-noise in the ultraviolet will primarily increase the star formation rate uncertainties.

\section{Results}
\label{sec:results}

\subsection{Global Properties}
\label{sec:global}
Figure~\ref{fig:sample} was briefly mentioned in \S~\ref{sec:sample} in the sample description.  This figure provides a selection of physical parameters and the ranges spanned by the sample of 32 galaxies studied here---proceeding from the top to bottom: stellar mass, star formation rate surface density, $g-r$ color, and stellar mass-weighted age \citep[see also Figure~2 and Figure~1, respectively, of][]{lopezfernandez18,casasola19}.  The values displayed in Figure~\ref{fig:sample} are either obtained directly from the observations in the case of $g-r$ color, or they are inferred from the SED fits.  These data are based on the integrated fluxes arising from within semi-major $\times$ semi-minor $1.5a_{25} \times 1.5b_{25}$ elliptical apertures (Table~\ref{tab:fluxes}).  Though the statistics suffer from small numbers when the galaxies are binned by morphology, the general trends are unsurprising.  The stellar mass is, on average, highest for the S0 and earlier-type spirals and generally decreases for later-type spirals and irregulars.  The S0 and early-type spirals are globally redder and older whereas the later-type spirals are bluer and younger.
 
\subsection{Spatially Resolved: Inner-disk vs Outer-Disk Properties}
\label{sec:resolved}
A primary goal of this effort is to quantify and compare the spatially resolved trends for the inner and outer disks.  Figure~\ref{fig:u7699.n4460.gr.age} shows two example trends for the six photometric annuli of UGC~7699 and NGC~4460: the $g-r$ colors and stellar (mass-weighted) ages.  The lines provided within the figure represent linear fits to the inner ($0<a<0.75a_{25}$) and outer ($0.75a_{25}<a<1.50a_{25}$) disks; the slope values are provided next to each fitted line in the figure.  For Scd galaxy UGC~7699, both the $g-r$ colors and fitted stellar ages decrease over the radial range $0<a<0.75a_{25}$ and increase over $0.75a_{25}<a<1.50a_{25}$, consistent with inside-out disk formation for the inner disk and a reversal for the outer disk.  The opposite trends are seen for S0 galaxy NGC~4460.  The sample-wide statistics for these two parameters are provided in Figure~\ref{fig:age_gr_vs_morph} as a function of morphology.  Negative (positive) gradients indicate younger (older) with radius (for reference, see the fitted lines and reported slopes in Figure~\ref{fig:u7699.n4460.gr.age}).  The general trend for the spiral galaxies in our 32 galaxy subset of the EDGES full sample is similar to that seen for UGC~7699 highlighted in Figure~\ref{fig:u7699.n4460.gr.age}: negative gradients for the inner disks and positive gradients for the outer disks.  The differences between the inner and outer disks for the later-type spiral galaxies can be more clearly seen in the histograms shown in Figure~\ref{fig:age_gr_hist}.  

The picture is less clear for S0 and irregular galaxies.  The $g-r$ color trends for S0s and irregulars span both positive and negative slopes for both inner and outer disks.  The average inner age gradient for irregulars is negative and the average outer age gradient is positive.  At face value this result contradicts the findings that irregular galaxies differ from spirals in that irregulars are more likely to exhibit features consistent with forming in an outside-in fashion \citep{gallart08,zhang12,meschin14,pan15,sacchi18}, where their shallow gravitational potentials lead to being more easily influenced by external factors like ram pressure stripping and internal factors such as feedback from stellar winds and supernovae.  However, there is significant overlap in the inner and outer age gradient distributions for the irregulars in Figure~\ref{fig:age_gr_vs_morph}.  Interestingly, for the S0s in our sample the age gradient is essentially reversed compared to the spirals: S0s appear to be increasingly older (younger) with increasing radius for the inner (outer) disk.  A complicating factor here is that four of the six S0 galaxies host Seyfert (NGC~4138 and NGC~5273) or LINER (NGC~4143 and NGC~4203) nuclei, and thus the observed optical colors and fitted stellar ages for their innermost annuli do not derive purely from star formation.  Only one spiral galaxy of the 16 studied here has such a nucleus; NGC~4102 is a LINER.  The CALIFA sample is only $\sim$6\% AGN \citep{walcher14}.

In their multi-wavelength study of 461 DustPedia galaxies, \cite{davies19} caution that a delayed star formation history model does not apply well to S0 galaxies.  Moreover, the formation histories of lenticulars in particular are thought to be quite diverse: in addition to a portion of S0s having been formed via mergers or the stripping of gas in clusters and groups, other S0 galaxies may have formed through internal secular processes and gas infall \citep{bellstedt17,elichemoral18,coccato19}.  The main conclusion of the \cite{davies19} study is that a delayed--$\tau$ model and closed-box chemical evolution can be appropriate for late-type spirals but earlier-type galaxies have more complicated evolutionary histories.  \cite{lopezfernandez18} similarly analyzed CALIFA star formation radial trends according to morphology, also using a delayed star formation history model, though their analysis only extends to twice the half-light optical radius (we typically probe to 4--6 half-light radii).  \cite{lopezfernandez18} find that E/S0 galaxies are, on average, older (smaller $\tau$) further from the nucleus whereas spiral galaxies are younger (larger $\tau$) farther out.  A different analysis of the CALIFA sample, focused on the inner half-light radius, shows that earlier-type galaxies have larger age gradients \citep{garciabenito17}, echoing our result displayed in Figure~\ref{fig:age_gr_vs_morph}.  In short, the 16 spiral galaxies in our sample differ from the six S0 and ten irregular galaxies in their radial gradients in optical color and characteristic stellar ages.  These gradient differences in their star-forming properties may reflect their disparate formation histories.  Though our project has been designed to probe more of the outer disk regions than IFU surveys like CALIFA, the statistical significance of our results are much less robust.  The \cite{lopezfernandez18} and \cite{garciabenito17} efforts, by comparison, respectively involve 48 and 78 S0 galaxies and 264 and 463 spiral galaxies.  Note that the \cite{lopezfernandez18} and \cite{garciabenito17} samples exclude Type~1 Seyferts and galaxies that exhibit strong merger or interaction features.

\section{Summary and Discussion}
\label{sec:summary}
The {\it Spitzer} EDGES survey provided extremely sensitive near-infrared maps, down to $\sim 0.4$~kJy or equivalently 29~AB~mag~arcsec$^{-2}$, for 92 nearby galaxies.  The near-infrared maps are large enough to reach out to five times the optical radius to enable detailed studies of the outermost extent of the stellar emission.  We provide here the global $ugr$ fluxes for a subset of 32 galaxies in the EDGES sample based on follow-up ground-based 60-minute integrations using the wide-field optical imager on the 2.3~m WIRO telescope.  We also report results on the galaxies' spatially resolved star formation histories.  The results are based on SED fits to the {\it Spitzer} near-infrared, WIRO optical, and ancillary ultraviolet and infrared data.  The CIGALE software package with a delayed--$\tau$ star formation history model is utilized for the SED fits; CIGALE is a software package that incorporates a balance between ultraviolet/optical stellar light that is absorbed by dust and an equal amount of energy that reappears at infrared wavelengths as dust emission.  

Analysis of the radial profiles provides evidence for a difference between the inner and outer gradients in the $g-r$ colors and characteristic stellar ages: extending out to three-quarters of the traditional optical radius, the 16 spiral galaxies show bluer colors and younger stellar ages with increasing radius, consistent with an inside-out disk formation scenario.  Outside this inner radial regime and extending to 50\% beyond the optical radius, the colors are redder and the stellar ages are older as a function of radius, consistent with either radial transport of inner stellar populations into the galaxy outskirts or a cumulative history of accretions/mergers that deposited previously external stellar systems into the galaxy haloes.  In both scenarios the outer stellar populations could be conspicuously redder and older: radial mixing moves some of the oldest stars in the galaxy centers toward the galaxy peripheries \citep{roskar08,sanchezblazquez09}, and simulations show that the bulk of a Milky Way-like galaxy's stellar halo at $z=0$ derives from the accumulated remnants of accreted satellites and merger tidal debris that occurred several Gyr ago \citep{bullock05,rodriguezgomez16}.  These remnants have passively evolved into an old red population residing in the present-day stellar halo.

The age and color gradients in the outer portions of the 10 irregular galaxies in our sample are similarly increasingly redder and older with radius.  The gradients for their inner portions span both positive and negative values, so overall these results do not suggest a coherent formation scenario for the irregular galaxies.  The six S0 galaxies have age and color gradients that are essentially the opposite that seen for the spiral galaxies: older and redder with radius for their inner portions, and younger and bluer for their outer disks.  In a panchromatic study of 461 DustPedia galaxies, \cite{davies19} recommend that a delayed star formation history is not applicable to earlier-type galaxies like S0s which do not lie along the galaxy main sequence and have more complicated formation histories.  Additional data and analysis for addtional EDGES targets are being pursued, including for the remaining six S0 and seven irregular galaxies, in order to see if these preliminary trends with morphology still hold.

%% file: acknowledgements.tex
\acknowledgements 
We thank M\'ed\'eric Boquien for help with the SED fitting.  This work is supported by the National Science Foundation under REU grant AST~1852289 and by NASA through an award issued by JPL/Caltech.  This work is based on observations made with the {\it Spitzer Space Telescope} and utilizes the NASA/IPAC Infrared Science Archive, both operated by JPL/Caltech under a contract with NASA.  {\em Herschel} is an ESA space observatory with science instruments provided by European-led Principal Investigator consortia and with important participation from NASA.  We gratefully acknowledge NASA's support for construction, operation, and science analysis for the GALEX mission, developed in cooperation with the Centre National d'Etudes Spatiales of France and the Korean Ministry of Science and Technology.  Funding for the Sloan Digital Sky Survey and SDSS-II has been provided by the Alfred P. Sloan Foundation, the Participating Institutions, the NSF, the U.S. Department of Energy, NASA, the Japanese Monbukagakusho, the Max Planck Society, and the Higher Education Funding Council for England.

%% file: tables.tex
\begin{deluxetable}{llclrcrcr}
\tablenum{1}
\def\d{$^\dagger$}
\def\p{$\pm$}
\tabletypesize{\scriptsize}
\tablecaption{Galaxy Sample}
\tablewidth{0pc}
\tablehead{
\colhead{Galaxy} &
\colhead{UGC} &
\colhead{$\alpha_0$~\&~$\delta_0$} &
\colhead{Optical} &
\colhead{$2a_{25}$} &
\colhead{$b_{25} \over a_{25}$} &
\colhead{c$z$} &
\colhead{$A_V$} &
\colhead{P.A.}
\\
\colhead{} &
\colhead{Number} &
\colhead{(J2000)} &
\colhead{Morphology} &
\colhead{(~$^\prime~$$^\prime~$)} &
\colhead{} &
\colhead{(km~s$^{-1}$)} &
\colhead{(mag)} &
\colhead{($\degr$)}
}
\startdata
\input table1.tex

\enddata
\tablecomments{\footnotesize The apertures used for the photometry have the centers and position angles (measured east of north) listed here, with ellipticities determined via $b_{25}/a_{25}$, where $2a_{25}$ and $2b_{25}$ are respectively the RC3 major axis and minor axis sizes of the $B$ band isophote defined at 25~mag~arcsec$^{-2}$ \citep{devaucouleurs91}.  All information is taken from the NASA/IPAC Extragalactic Database (NED) including the foreground Milky Extinction.}
\label{tab:sample}
\end{deluxetable}

\begin{deluxetable}{lll}
\tablenum{2}
\tabletypesize{\scriptsize}
\tablecaption{Fit Parameters}
\tablewidth{0pc}
\tablehead{
\colhead{Parameter} &
\colhead{Notation} &
\colhead{Allowed Values} 
}
\startdata
\input table2.tex
\enddata
\label{tab:parameters}
\end{deluxetable}

\begin{deluxetable}{lllllllll}
\tablenum{3}
\def\a{\tablenotemark{a}}
\def\b{\tablenotemark{b}}
\def\c{\tablenotemark{c}}
\def\d{\tablenotemark{d}}

\def\p{$\pm$}
\tabletypesize{\scriptsize}
\tablecaption{Integrated Fluxes}
\tablewidth{0pc}
\tablehead{
\colhead{Galaxy} &
\colhead{\it GALEX} &
\colhead{\it GALEX} &
\colhead{WIRO} &
\colhead{WIRO} &
\colhead{WIRO} &
\colhead{\it Spitzer} &
\colhead{WISE} &
\colhead{\it Spitzer} 
\\
\colhead{} &
\colhead{FUV} &
\colhead{NUV} &
\colhead{$u$} &
\colhead{$g$} &
\colhead{$r$} &
\colhead{3.6\m} &
\colhead{12\m} &
\colhead{24\m} 
}
\startdata
\input table3.tex

\enddata
\tablecomments{Total fluxes (in mJy) are derived using semi-major $\times$ semi-minor elliptical apertures of $1.5a_{25} \times 1.5b_{25}$.  The compact table entry format TUV$\pm$WXYEZ implies (T.UV$\pm$W.XY)$\times10^{\rm Z}$.  All fluxes were corrected for Galactic extinction \citep{schlafly11} assuming $A_V/E(B-V)\approx3.1$ and the reddening curve of \cite{draine03}.  The uncertainties include both statistical and systematic effects.}
\tablenotetext{a} {{\it WISE} 22\m}
\tablenotetext{b} {{\it Herschel} 70\m\ \citep{melendez14}}
\tablenotetext{c} {{\it Spitzer} 8\m}
\tablenotetext{d} {Ultraviolet emission extends beyond the aperture \citep{thilker07}.}
\label{tab:fluxes}
\end{deluxetable}


%% file: table1.tex
NGC4085&UGC07075&120522.7$+$502111&SAB(s)c?    &169.1&0.28& 746&0.050& 78\\
NGC4088&UGC07081&120534.2$+$503221&SAB(rs)bc   &345.3&0.39& 757&0.054& 43\\  
NGC4096&UGC07090&120601.1$+$472842&SAB(rs)c    &396.4&0.27& 566&0.050& 20\\  
NGC4102&UGC07096&120623.0$+$524240&SAB(s)b?    &181.2&0.57& 846&0.055& 38\\ 
NGC4138&UGC07139&120929.8$+$434107&SA0$^+$(r)  &154.2&0.66& 888&0.039&150\\ 
NGC4143&UGC07142&120936.0$+$423203&SAB0$^0$(s) &137.5&0.63& 946&0.035&144\\ 
NGC4203&UGC07256&121505.0$+$331150&SAB0$^-$?   &203.3&0.93&1086&0.033& 10\\ 
NGC4214&UGC07278&121539.2$+$361937&IAB(s)m     &510.7&0.78& 291&0.060&  0\\ 
NGC4220&UGC07290&121611.7$+$475300&SA0         &233.4&0.35& 914&0.049&140\\
\nodata&UGC07301&121642.1$+$460444&Sd          &109.2&0.13& 690&0.030& 82\\
NGC4242&UGC07323&121730.2$+$453709&SAB(s)dm    &300.7&0.76& 506&0.033& 27\\
DDO120 &UGC07408&122115.3$+$454850&IAm         &157.8&0.46& 462&0.032&100\\
NGC4369&UGC07489&122436.2$+$392259&(R)SA(rs)a  &125.4&0.98&1045&0.070&  0\\ 
DDO125 &UGC07577&122740.9$+$432944&Im          &255.9&0.56& 195&0.056&130\\ 
DDO129 &UGC07608&122844.2$+$431327&Im          &203.3&0.98& 538&0.047&  0\\ 
NGC4460&UGC07611&122845.5$+$445151&SB0$^+$(s)? &238.9&0.29& 490&0.052& 40\\
\nodata&UGC07639&122953.4$+$473152&Im          &137.5&0.71& 382&0.032&153\\
NGC4485&UGC07648&123031.1$+$414204&IB(s)m pec  &137.5&0.71& 493&0.059&  4\\
NGC4490&UGC07651&123036.2$+$413838&SB(s)d pec  &378.6&0.49& 565&0.060&121\\
\nodata&UGC07699&123248.0$+$373718&SBcd?       &228.1&0.27& 496&0.033& 32\\
NGC4618&UGC07853&124132.8$+$410903&SB(rs)m     &250.1&0.81& 544&0.058& 27\\
NGC4625&UGC07861&124152.7$+$411626&SAB(rs)m pec&131.3&0.87& 621&0.050&133\\
IC3687 &UGC07866&124215.1$+$383012&IAB(s)m     &203.3&0.89& 354&0.055&  0\\ 
NGC4707&UGC07971&124822.9$+$510953&Sm?         &134.3&0.93& 468&0.030& 23\\
IC4182 &UGC08188&130550.9$+$373601&SA(s)m      &361.5&0.91& 321&0.038&100\\ 
DDO166 &UGC08303&131317.6$+$361303&IAB(s)m     &134.3&0.85& 944&0.049&177\\
DDO168 &UGC08320&131427.9$+$455509&IBm         &217.8&0.38& 192&0.042&150\\
NGC5055&UGC08334&131549.3$+$420145&SA(rs)bc    &755.4&0.57& 484&0.048& 94\\
NGC5229&UGC08550&133402.8$+$475456&SB(s)d?     &198.7&0.17& 364&0.049&167\\
NGC5273&UGC08675&134208.3$+$353915&SA0(s)      &165.3&0.91&1085&0.028&  4\\ 
NGC5523&UGC09119&141452.3$+$251903&SA(s)cd?    &274.3&0.28&1039&0.052& 92\\
NGC5608&UGC09219&142317.9$+$414633&Im?         &157.8&0.51& 663&0.026& 95\\

%% file: table2.tex
Metallicity	& $Z$	& 0.008, 0.02, 0.05 \\
Stellar library &       & Bruzual \& Charlot (2003)\\
Initial Mass Function		&	& Chabrier (2003)\\ 
Color excess: young stars &$E(B-V)_*^{\rm y}$ & 0.0, 0.025, 0.05, 0.1, 0.15, 0.2, 0.25, 0.3, 0.4	 \\
Color excess: old stars &$E(B-V)_*^{\rm o}$ & 0.44$E(B-V)_*^{\rm y}$	 \\
Dust emission template & $\alpha$	& 0.5, 1.0, 1.25, 1.50, 1.75, 2.0, 2.25, 2.50, 3.00\\
Slope of power law that modifies attenuation curve&$\delta$ & $-$0.5, $-$0.4, $-$0.3, $-$0.2, $-$0.1, 0 \\
\hline
\multicolumn{3}{c}{Delayed Star Formation History} \\
\hline
SFR $e$-folding time (Gyr)	& $\tau$ & 0.5, 1, 1.5, 2, 2.5, 3, 3.5, 4, 4.5, 5, 5.5, 6, 6.5, 7, 7.5, 8, 10\\
Age of oldest stars (Gyr ago) & $t_0$ & 11\\	
\hline

%% file: table3.tex
NGC4220  &494\p025E$-$1&142\p007E$+$0&154\p007E$+$1&712\p028E$+$1&143\p005E$+$2&196\p009E$+$2&141\p014E$+$2  &147\p017E$+$2\\
UGC7301  &415\p020E$-$1&567\p028E$-$1&129\p005E$+$0&344\p013E$+$0&518\p021E$+$0&311\p015E$+$0&\nodata        &\nodata      \\
NGC4242  &891\p044E$+$0&140\p007E$+$1&422\p050E$+$1&123\p005E$+$2&159\p007E$+$2&120\p006E$+$2&847\p106E$+$1  &109\p017E$+$2\\
NGC4485  &124\p006E$+$1&163\p008E$+$1&284\p011E$+$1&571\p023E$+$1&979\p040E$+$1&419\p021E$+$1&961\p097E$+$1  &201\p015E$+$2\\
NGC4490  &550\p030E$+$1&875\p043E$+$1&203\p008E$+$2&449\p018E$+$2&807\p032E$+$2&519\p026E$+$2&152\p015E$+$3  &428\p030E$+$3\\
NGC4618  &261\p013E$+$1&333\p016E$+$1&632\p034E$+$1&158\p006E$+$2&213\p009E$+$2&167\p008E$+$2&258\p026E$+$2  &397\p030E$+$2\\
NGC4625\d&409\p025E$+$0&594\p029E$+$0&135\p013E$+$1&344\p014E$+$1&535\p026E$+$1&493\p024E$+$1&101\p010E$+$2  &127\p009E$+$2\\
NGC4707  &329\p017E$+$0&374\p021E$+$0&599\p036E$+$0&136\p008E$+$1&187\p010E$+$1&115\p006E$+$1&\nodata        &\nodata      \\
UGC8303  &351\p018E$+$0&423\p021E$+$0&691\p059E$+$0&174\p012E$+$1&219\p009E$+$1&137\p007E$+$1&134\p024E$+$1  &511\p075E$+$1\a\\
UGC8320  &533\p027E$+$0&707\p036E$+$0&\nodata      &281\p011E$+$1&364\p016E$+$1&188\p009E$+$1&272\p089E$+$0  &112\p022E$+$1\\
NGC5055\d&364\p018E$+$1&663\p036E$+$1&\nodata      &108\p004E$+$3&187\p007E$+$3&254\p012E$+$3&498\p050E$+$3  &574\p040E$+$3\\
NGC5229  &182\p009E$+$0&251\p012E$+$0&536\p027E$+$0&122\p005E$+$1&184\p007E$+$1&117\p006E$+$1&444\p115E$+$0  &131\p016E$+$1\\
NGC5273  &223\p015E$-$1&104\p012E$+$0&155\p007E$+$1&648\p026E$+$1&127\p005E$+$2&124\p006E$+$2&416\p054E$+$1  &294\p097E$+$3\b\\
NGC5523  &402\p021E$+$0&587\p030E$+$0&149\p007E$+$1&395\p017E$+$1&616\p026E$+$1&542\p027E$+$1&749\p075E$+$1  &136\p015E$+$2\a\\
NGC5608  &321\p016E$+$0&387\p019E$+$0&660\p034E$+$0&147\p006E$+$1&200\p008E$+$1&109\p005E$+$1&295\p097E$+$0  &607\p200E$+$0\a\\
NGC4085  &155\p007E$+$0&282\p015E$+$0&104\p004E$+$1&319\p013E$+$1&570\p024E$+$1&977\p049E$+$1&313\p031E$+$2  &577\p041E$+$2\a\\
NGC4088  &102\p005E$+$1&184\p009E$+$1&577\p039E$+$1&177\p007E$+$2&303\p013E$+$2&459\p023E$+$2&162\p016E$+$3  &327\p023E$+$3\a\\
NGC4096  &732\p036E$+$0&126\p007E$+$1&438\p107E$+$1&136\p006E$+$2&228\p010E$+$2&300\p015E$+$2&109\p011E$+$3\c&846\p064E$+$2\\
NGC4102  &203\p025E$+$0&433\p038E$+$0&372\p016E$+$1&155\p006E$+$2&309\p012E$+$2&340\p017E$+$2&117\p011E$+$3  &584\p056E$+$4\b\\
NGC4138  &231\p011E$+$0&364\p018E$+$0&200\p016E$+$1&820\p036E$+$1&154\p006E$+$2&192\p009E$+$2&167\p016E$+$2  &310\p102E$+$3\b\\
NGC4143  &443\p022E$-$1&120\p014E$+$0&165\p006E$+$1&875\p035E$+$1&176\p007E$+$2&239\p011E$+$2&471\p053E$+$1  &514\p096E$+$1\a\\
NGC4203  &619\p033E$-$1&204\p011E$+$0&231\p076E$+$1&147\p011E$+$2&244\p047E$+$2&369\p018E$+$2&104\p015E$+$2  &722\p115E$+$1\\
NGC4214  &836\p041E$+$1&107\p005E$+$2&209\p011E$+$2&407\p017E$+$2&594\p028E$+$2&346\p017E$+$2&672\p089E$+$2\c&200\p015E$+$3\\
UGC7408  &120\p006E$+$0&210\p011E$+$0&484\p023E$+$0&138\p006E$+$1&178\p009E$+$1&108\p006E$+$1&000\p000E$-$2  &000\p000E$-$2\\
NGC4369  &431\p021E$+$0&770\p038E$+$0&219\p017E$+$1&634\p025E$+$1&108\p004E$+$2&109\p005E$+$2&241\p024E$+$2  &632\p047E$+$2\a\\
UGC7577  &377\p018E$+$0&546\p027E$+$0&111\p025E$+$1&315\p030E$+$1&482\p047E$+$1&263\p019E$+$1&322\p106E$+$0\c&243\p080E$+$0\\
UGC7608  &409\p020E$+$0&462\p044E$+$0&669\p041E$+$0&143\p007E$+$1&168\p026E$+$1&117\p014E$+$1&282\p093E$+$0\c&205\p067E$+$1\\
NGC4460  &215\p010E$+$0&443\p023E$+$0&205\p008E$+$1&688\p027E$+$1&129\p005E$+$2&860\p043E$+$1&154\p015E$+$2\c&300\p021E$+$2\\
UGC7639  &114\p005E$+$0&184\p010E$+$0&382\p024E$+$0&121\p005E$+$1&187\p009E$+$1&113\p010E$+$1&301\p099E$+$0\c&351\p116E$+$0\\
UGC7699  &411\p020E$+$0&585\p030E$+$0&111\p004E$+$1&269\p010E$+$1&349\p014E$+$1&231\p011E$+$1&236\p033E$+$1\c&348\p038E$+$1\\
IC~3687  &453\p022E$+$0&532\p026E$+$0&757\p044E$+$0&170\p011E$+$1&218\p011E$+$1&124\p010E$+$1&243\p080E$+$0\c&666\p220E$+$0\\
IC~4182  &148\p007E$+$1&195\p011E$+$1&327\p107E$+$1&902\p046E$+$1&125\p005E$+$2&778\p042E$+$1&473\p156E$+$1\c&713\p235E$+$1\\

%% file: figs.tex

\begin{figure}
 \plotone{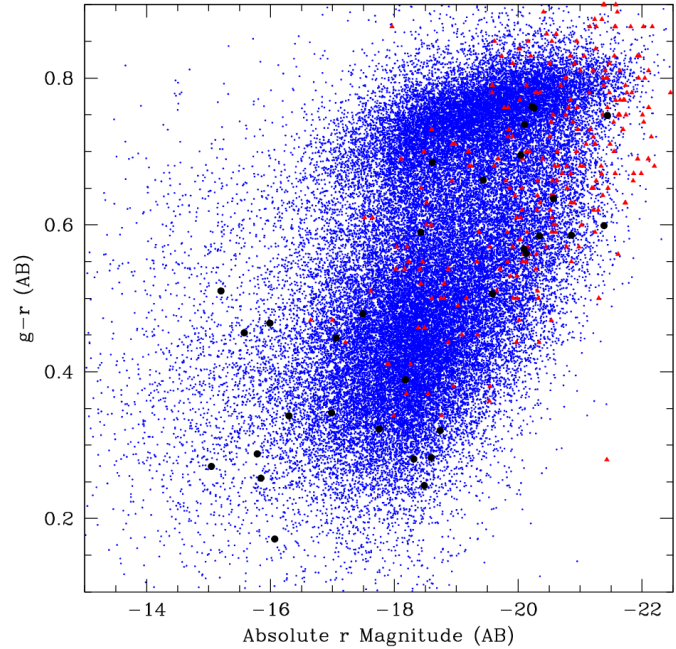}
 \caption{Comparison of global $g - r$ colors and absolute $r$ magnitudes for the EDGES subsample studied here (large black circles), the spiral galaxies in the CALIFA survey (red triangles are spirals from the DR2 release; \citealt{garciabenito15,rodriguezbaras18}), and the \SDSS\ low redshift sample (small blue dots; $10<d<150~{\rm Mpc}~h^{-1}$; \citealt{blanton05}).  The values are corrected for foreground Milky Way attenuation.}
 \label{fig:color_luminosity}
\end{figure}

\begin{figure}
 \plotone{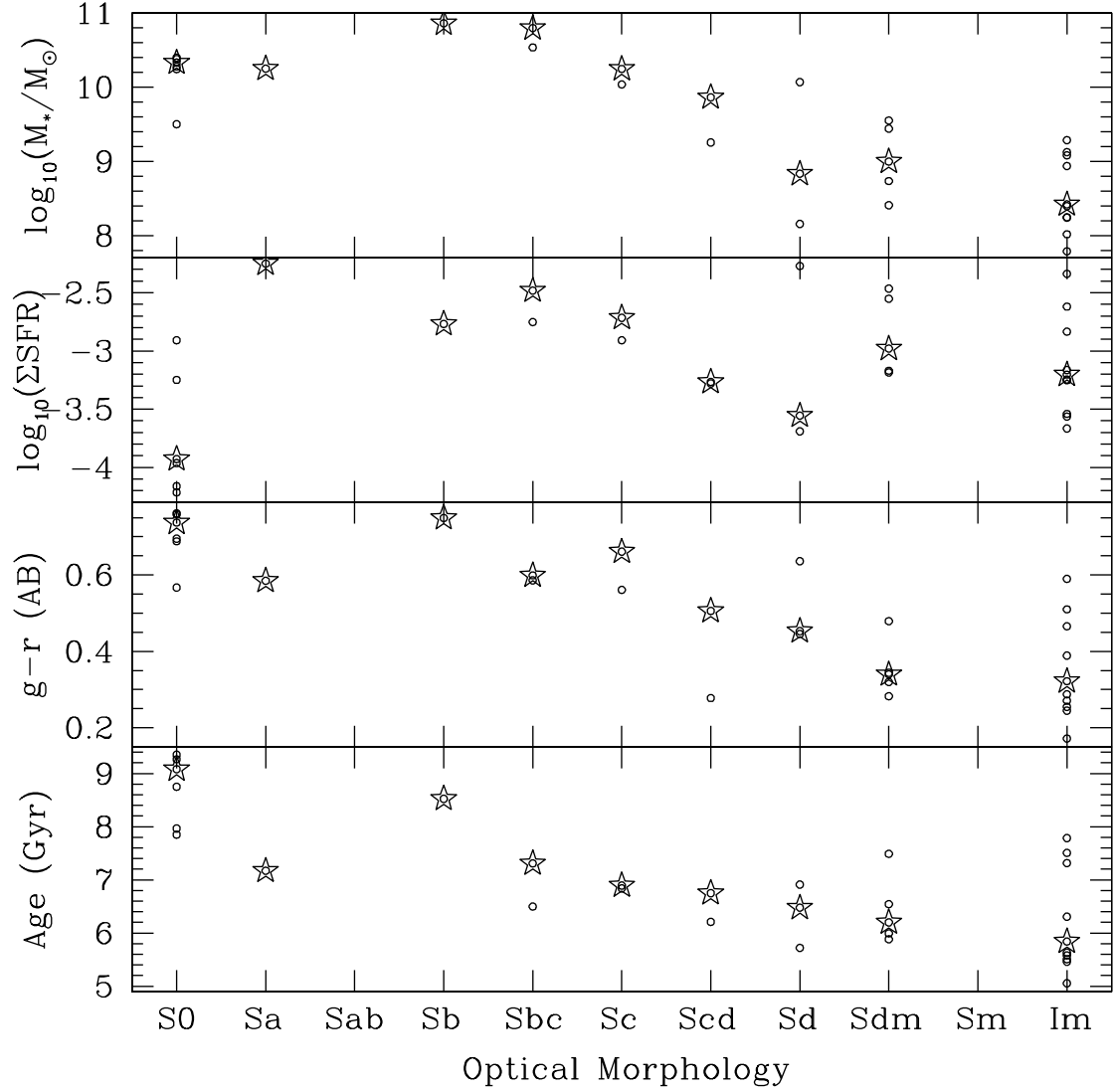}
 \caption{Compilation of sample characteristics derived from our global photometry and SED fitting.  Top row: stellar mass.  Second row: star formation rate surface density in $M_\odot~{\rm yr}^{-1}~{\rm kpc}^{-2}$.  Third row: global $g-r$ color.  Bottom row: stellar mass-weighted age.  The large stars indicate median values.}
 \label{fig:sample}
\end{figure}

\begin{figure}
 \includegraphics[angle=270,width=8in]{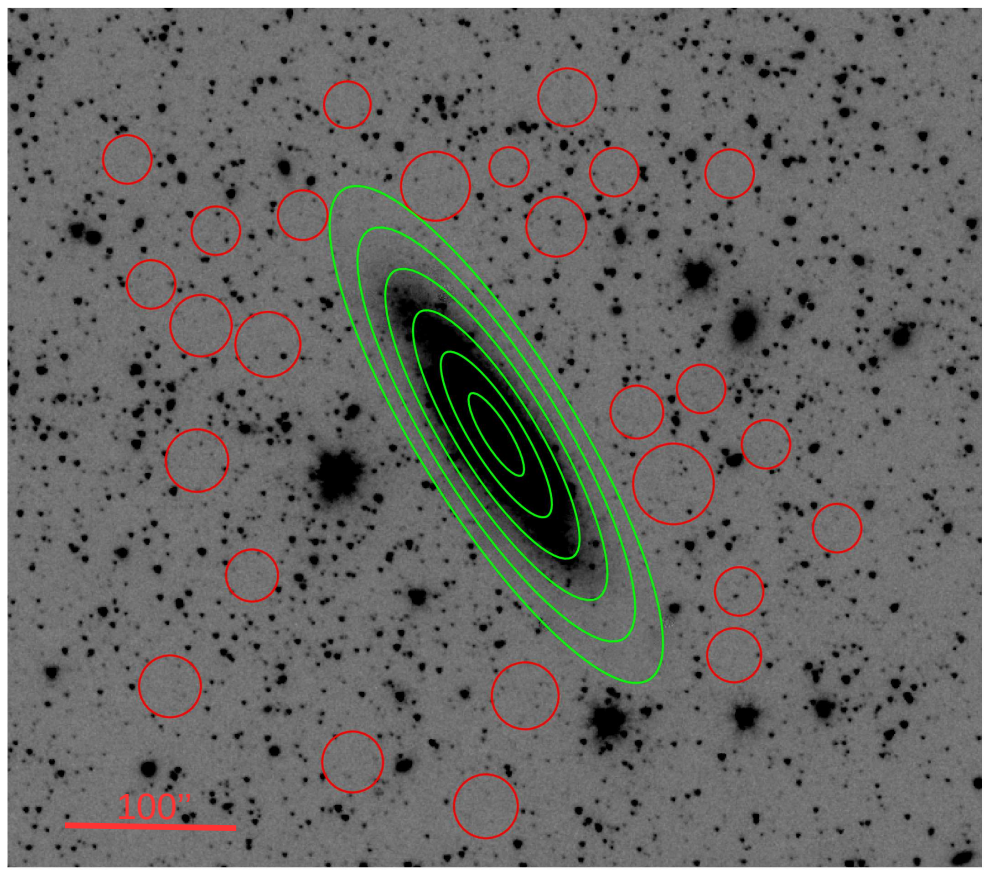}
 \caption{A 9\farcm6$\times$8\farcm5 portion of the {\it Spitzer} 3.6\m\ mosaic of UGC~07699.  The ellipses demonstrate the annular regions for extracting photometry and the red circles show the sky apertures.  North is up, East is to the left.  The angular scale is provided in the lower left}.
 \label{fig:aps}
\end{figure}

\begin{figure}
 \plotone{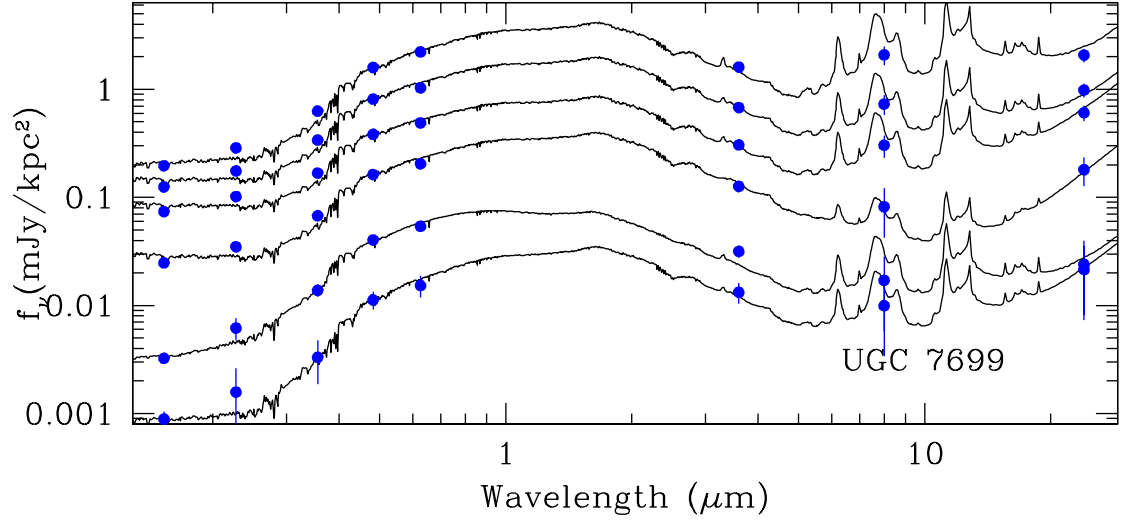}
 \caption{SEDs for the six annular regions of UGC~07699.  The blue dots indicate the measured surface brightnesses and the black curves show the best-matched stellar+dust SEDs assuming a delayed star formation history.  The $\chi^2_{\rm reduced}$ values for the fits, proceeding from the innermost annulus to the outermost annulus, are 1.9, 1.2, 1.1, 1.4, 1.5, and 0.5.}
 \label{fig:sed}
\end{figure}

\begin{figure}
 \plotone{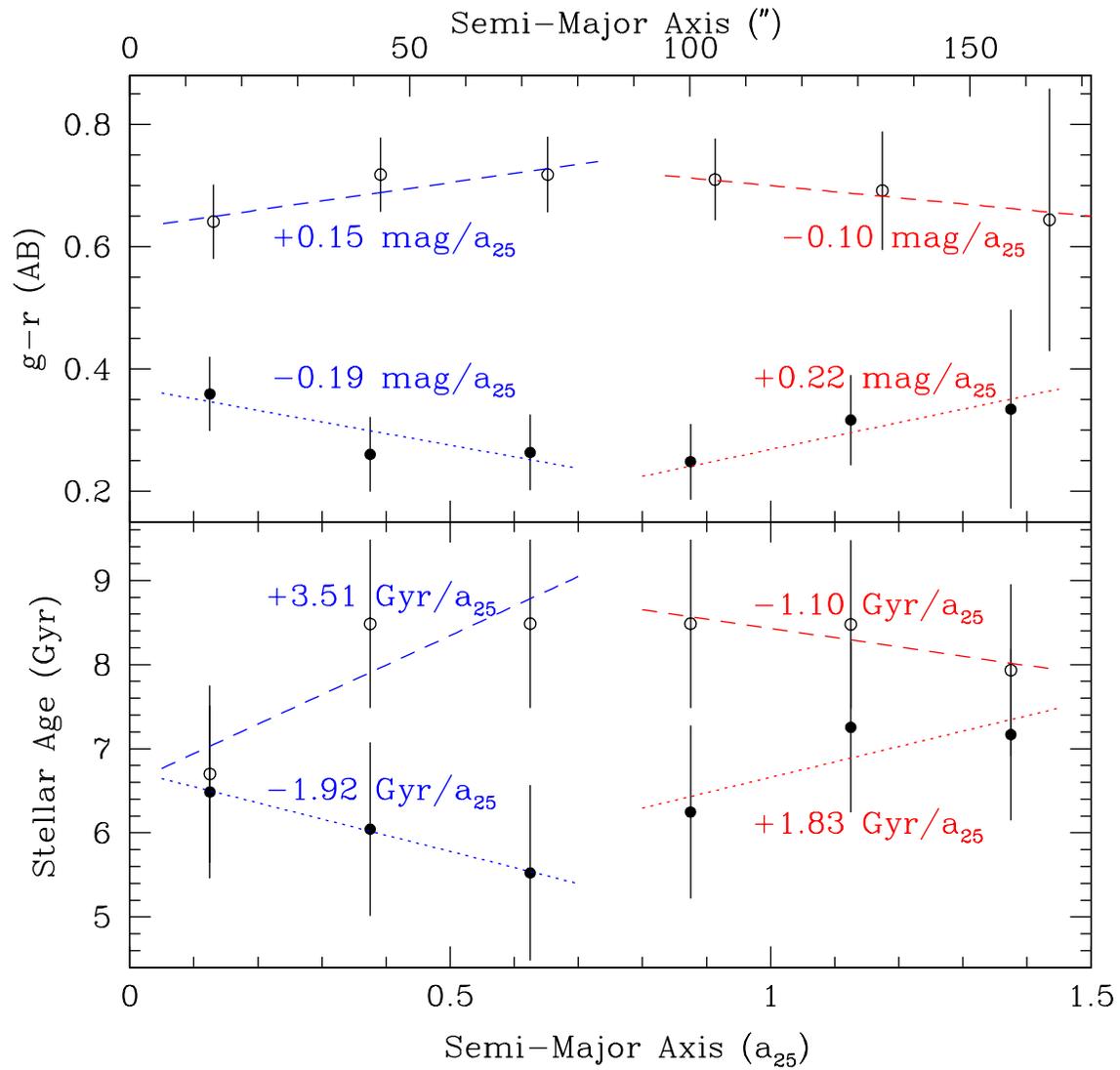}
 \caption{Examples of $g-r$ and stellar mass-weighted age radial profiles, for Scd UGC~07699 (filled circles and dotted lines) and S0 NGC~4460 (open circles and dashed lines).  Dotted lines indicate linear fits to the inner and outer portions of the radial profiles.  The slopes of the linear fits are also provided.}
 \label{fig:u7699.n4460.gr.age}
\end{figure}

\begin{figure}
 \plotone{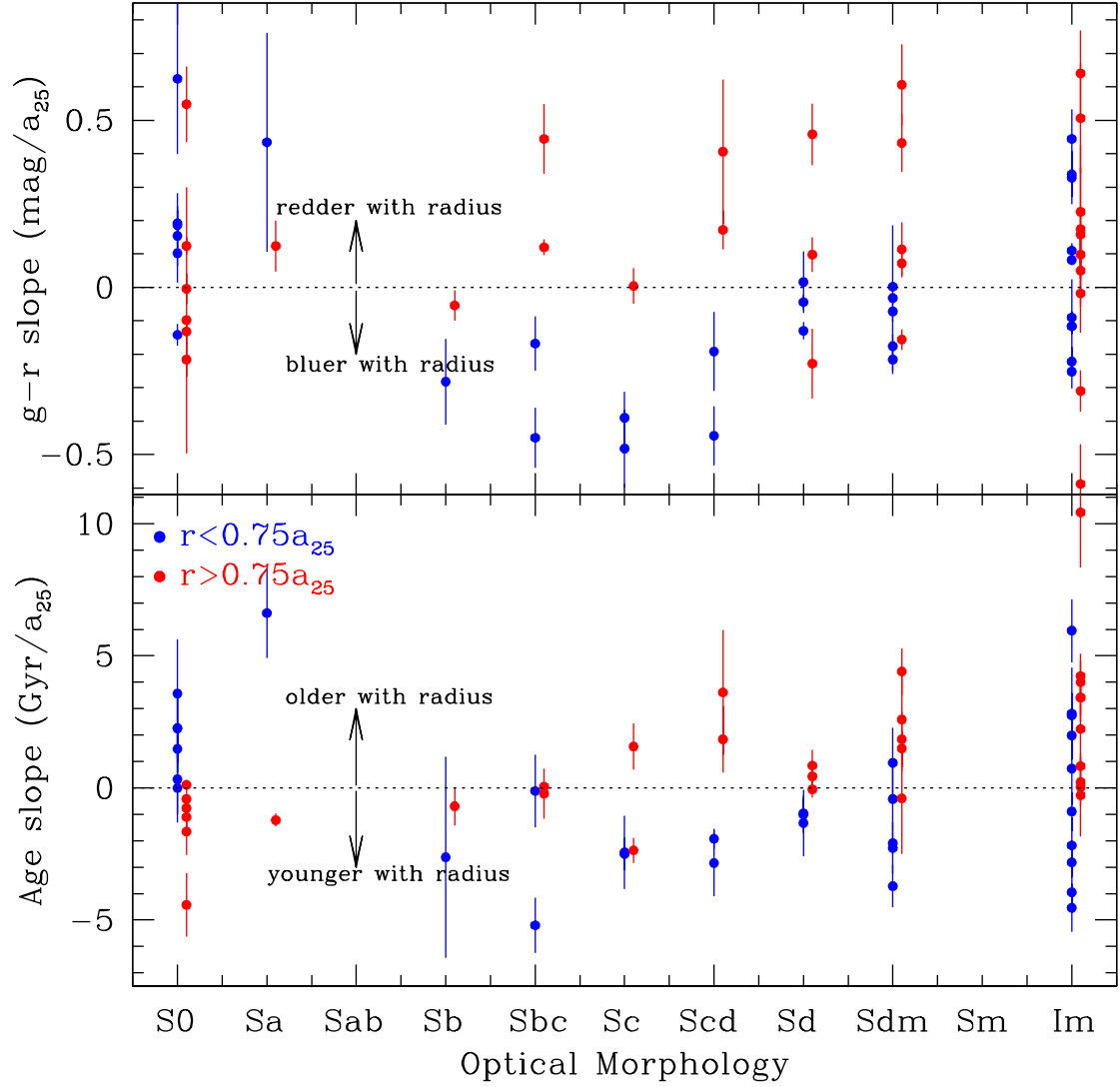}
 \caption{The $g-r$ and stellar age radial slopes as a function of optical morphology (see also Figure~\ref{fig:u7699.n4460.gr.age} for example slopes).  The data are color-coded according to radial extent: blue (red) indicates radii less (greater) than 0.75$a_{25}$.}
 \label{fig:age_gr_vs_morph}
\end{figure}

\begin{figure}
 \plotone{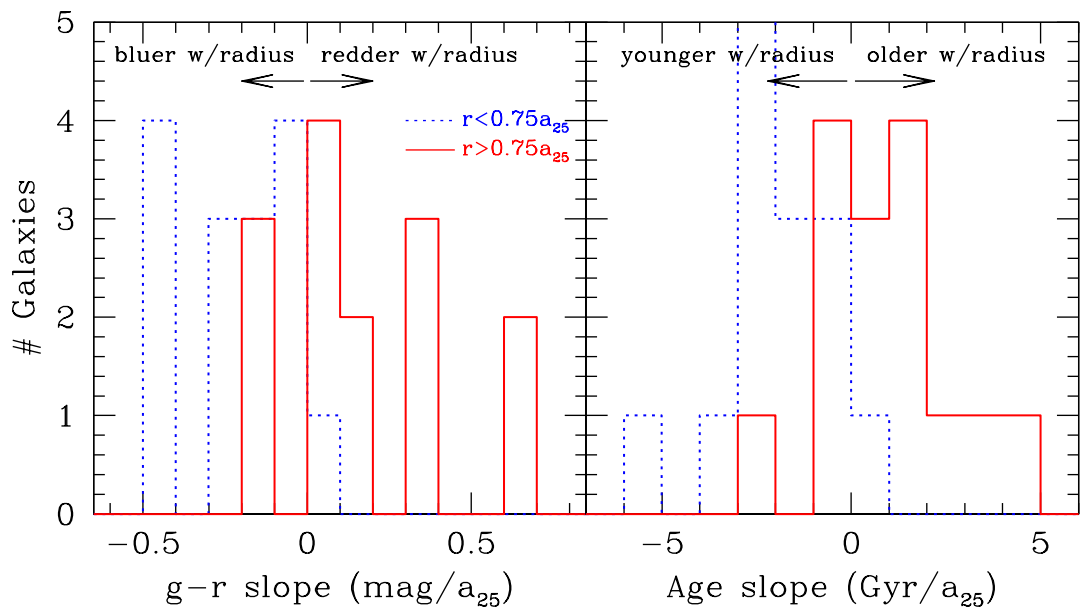}
 \caption{Histograms of the $g-r$ and stellar age radial slopes, differentiated according to inner (blue dotted) and outer (red solid) radial extents.  Excludes S0/Sa and irregular galaxies (and there are no ellipticals in the sample).}
 \label{fig:age_gr_hist}
\end{figure}